\documentclass[letterpaper, 10pt, conference]{ieeeconf}      

\IEEEoverridecommandlockouts                              

\usepackage{graphics} 
\usepackage{epsfig} 
\usepackage{mathptmx} 
\usepackage{times} 
\usepackage{amsmath} 
\usepackage{amssymb}  
\usepackage{subcaption}
\usepackage{mathtools}
\usepackage{mathrsfs}
\usepackage{placeins}
\usepackage{float}
\usepackage{multicol}
\usepackage{xspace}
\usepackage{mathpazo}
\usepackage[capitalise]{cleveref}
\usepackage{siunitx}
\newcommand{\MATLAB}{\textsc{Matlab}\xspace}
\usepackage{xcolor}
\usepackage{epstopdf} 
\usepackage{multirow}
\usepackage[utf8x]{inputenc}
\title{\LARGE \bf
Self-adaptive Torque Vectoring Controller Using Reinforcement Learning
}

\author{Shayan Taherian, Sampo Kuutti, Marco Visca and Saber Fallah 
\thanks{Shayan Taherian, Sampo Kuutti, Marco Visca and Saber Fallah are with CAV-Lab in Department of Mechanical Engineering,
        University of Surrey, Guildford, UK
        {\tt\small s.fallah@surrey.ac.uk}}%
}

\begin{document}

\maketitle
\thispagestyle{empty}
\pagestyle{empty}

\begin{abstract}
Continuous direct yaw moment control systems such as torque-vectoring controller are an essential part for vehicle stabilization.  This controller has been extensively researched with the central objective of maintaining the vehicle stability by providing consistent stable cornering response. The ability of careful tuning of the parameters in a torque-vectoring controller can significantly enhance vehicle's performance and stability. However, without any re-tuning of the parameters, especially in extreme driving conditions e.g. low friction surface or high velocity, the vehicle fails to maintain the stability. In this paper, the utility of Reinforcement Learning (RL) based on Deep Deterministic Policy Gradient (DDPG) as a parameter tuning algorithm for torque-vectoring controller is presented. It is shown that, torque-vectoring controller with parameter tuning via reinforcement learning performs well on a range of different driving environment e.g., wide range of friction conditions and different velocities, which highlight the advantages of reinforcement learning as an adaptive algorithm for parameter tuning. Moreover, the robustness of DDPG algorithm are validated under scenarios which are beyond the training environment of the reinforcement learning algorithm. The simulation has been carried out using a four wheels vehicle model with nonlinear tire characteristics. We compare our DDPG based parameter tuning against a genetic algorithm and a conventional trial-and-error tunning of the torque vectoring controller, and the results demonstrated that the reinforcement learning based parameter tuning significantly improves the stability of the vehicle.
 
\end{abstract}

\section{INTRODUCTION}

Active stability technologies play a major role in modern vehicle dynamic systems in order to enhance the lateral vehicle stability and reduction in fatal accidents. These technologies such as torque-vectoring controllers (i.e., the control of the traction and braking torque of each wheel), can effectively enhance the vehicle handling performance in extreme manoeuvres \cite{Lu2016}\cite{Wang2018}. Torque-vectoring controller has a prominent advantage of stabilizing a vehicle through choosing appropriate control parameters. However, if the parameters cannot be adapted to the environmental changes, then control design would be cumbersome and will not guarantee good robustness. 
In order to solve this problem, the adaptive control design has received wide attention. Some methods of tuning parameters have been proposed such as fuzzy logic method \cite{Fu2016}\cite{VanKien2019}, which have optimization problem of parameters and requires more knowledge of the environment. Another method is evolutionary algorithm \cite{Cao2020}\cite{Rajinikanth2012} such as genetic algorithm, which is a strong tool for tuning the control parameter with less computational requirements for prior knowledge, however depending on the complexity of the problem the calculation speed will be slow. Others have used neural networks to learn the optimal parameters for each state \cite{Fei2018}\cite{Chen2004}, typically using supervised learning to train the neural network. However, the downside of supervised learning methods is that during training, the ground truth (i.e. correct prediction for given inputs) for each prediction is required to measure the error in the network's predictions. Therefore, supervised learning is poorly suited to problems where the ground truth is not readily available or is difficult to estimate. On the other hand, reinforcement learning is a technique inspired by human and animal learning, where the neural network is trained through a trial-and-error mechanism, without requiring any ground truth labels. Therefore, reinforcement learning is better suited for this type of problem where the correct predictions are not known, as it enables the network to learn the optimal policy effectively through interaction with its environment. The success of reinforcement learning \cite{Silver2016}\cite{Silver2017} was perceived as a breakthrough in the field of artificial intelligence. This algorithm attempts to learn the optimal control policy through interactions between the system and its environment without the need of prior knowledge of an environment or the system model. In this algorithm, an agent takes an action based on environment state and consequently receives a reward. Reinforcement learning algorithms are generally divided into three groups \cite{Konda2003}\cite{Hsu2019}: value based, policy gradient and actor-critic methods. In this paper, DDPG algorithm \cite{Lillicrap2016} is used, which is in the family of actor-critic group algorithm, to tune the parameters of the torque-vectoring controller. DDPG agent is an actor-critic reinforcement learning that computes optimal policy that maximizes the long-term reward. The actor is a network attempting to execute the best action given the current state. The critic estimates the maximum value function of given state (i.e approximate maximizer). Then uses the reward from environment to determines the accuracy of the prediction value. In \cite{Wang2019}, authors proposed DDPG as a reinforcement learning algorithm for tuning the sliding mode controller parameters for autonomous underwater vehicle. Results demonstrated that DDPG achieved a good performance in terms of stability, fast convergence and low chattering. In \cite{WANG2007}, reinforcement learning was used to tune PID controller parameters where the auto tuning strategy was through actor-critic reinforcement learning method resulting in stable tracking performance of the system. However, the generalization of the trained models for the scenarios beyond the training environment were not validated. Therefore the performance of the algorithms in new environments can not be guaranteed.\\
 In this paper, reinforcement learning algorithm is incorporated into the torque-vectoring controller to adjust the control input weights in order to maintain the stability of the vehicle. The contribution in this paper lies into the utilization of DDPG algorithm as an intelligent tuning strategy for the torque-vectoring controller in a real application for different scenarios. It has been shown that proper tuning of the control input weights results in improvement in handling and stability of the vehicle. In order to validate the effectiveness of the DDPG algorithm, simulation environment has been set under different velocities ranging from 80-130  $\SI{}{\kilo\metre\per\hour}$ and different friction surface ranging from 0.4-0.6. The performance is then compared with a conventional trial-and-error approach (which will be referred as manual tuning for the rest of the paper), and a self-tuning genetic algorithm, to show the enhanced effectiveness of DDPG as a tuning strategy for active yaw control systems. Moreover, to validate the generalization capability of the trained model, simulation has been deployed in the environment that DDPG was not trained in, to ensure the robustness of the algorithm under unknown operating conditions. The simulation results have been conducted using a four wheels vehicle model \cite{Falcone2007a} with nonlinear tire characteristics in \MATLAB/Simulink environment. Moreover, for solving the on-line tuning algorithm, \MATLAB reinforcement learning toolbox is used for parameter tuning of the torque-vectoring controller.  
The remainder of the paper is structured as follows: \cref{vehicle-model} describes the vehicle model used in this paper. \cref{torque} introduces the torque vectoring algorithm used for stabilization purposes in this paper. \cref{RL} presents the reinforcement learning algorithm used for tuning parameters of the torque vectoring control. \cref{S-results} demonstrates the simulation results of the trained algorithm, and finally concluding remarks are presented in \cref{conclusion}.
\section{Vehicle model} 
\label{vehicle-model}
In this paper, vehicle simulation model is the six state nonlinear four wheel vehicle model presented in \cite{Falcone2007a}. The model states are $[\dot{x}, \dot{y}, \psi, \dot{\psi}, X, Y]^T$ where $\dot{x}$ and $\dot{y}$ represents the vehicle longitudinal and lateral velocities, respectively, in the body frame and $\psi$, $\dot{\psi}$, $X$ and $Y$ denotes yaw angle, yaw rate, longitudinal and lateral coordinates in the inertial frame. To describe the vehicle model, normal forces are assumed to be constant and roll angle is neglected in the formulation. The tire model used in this paper is described by pacejka model \cite{Pacejka2012a}. This is an empirical nonlinear tire model which is able to closely capture the tire dynamics.

\section{Torque vectoring control}
\label{torque}
\subsection{System Architecture}
\begin{figure*}[t!]
	\begin{multicols}{2}
		\includegraphics[width=1.9\linewidth]{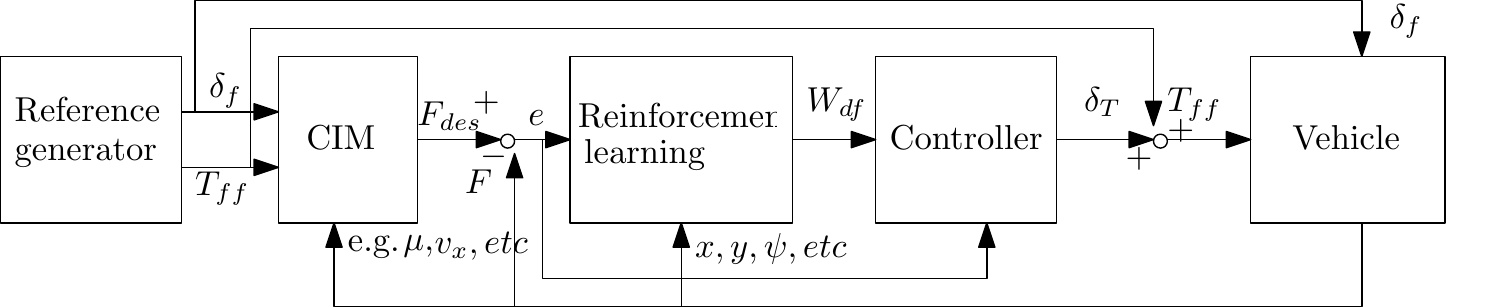}
		\hspace{16cm}
	\end{multicols}
	\caption{Adaptive torque vectoring controller based on reinforcement learning}
	\label{structure}
\end{figure*}

The overall architecture of the proposed control framework is shown in \cref{structure}. In this structure, the command interpreter module (CIM) generates desired forces at the centre of gravity (C.G) based on inputs from the reference generation block ($\delta_f$ and $T_{ff}$) in order to interpret the driver's intention for the vehicle motion. CIM continuously monitors the inputs as well as the states of the vehicle to provide accurate desired values to the torque vectoring controller to keep the vehicle within the stable region. It is noted that CIM generats the desired C.G forces with assumption of normal driving conditions (e.g. high surface fricition condition). For the brevity of the paper, the calculation of CIM block is referred to \cite{Fallah2013}. The inputs to the reinforcement learning are the errors between the actual and desired forces followed by the actual states of the vehicle. Moreover, the generated error between the actual and reference signal as well as the actions from reinforcement learning ($W_{df}$) are fed to the controller in order to generate the adjustment torque required to assist the driver for improving handling and stability of the vehicle. 
\subsection{Controller}
\label{control}
\begin{figure}[t!]
 		\hspace*{-0.2in}
 		\centering
 		\includegraphics[width=0.55\linewidth]{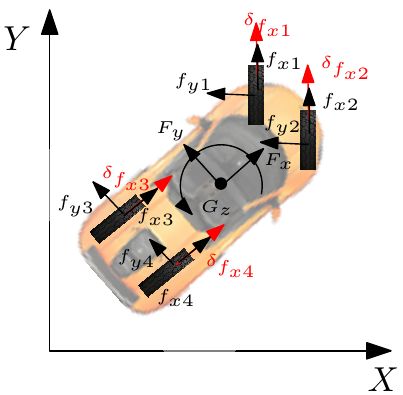}
 		\caption {Force convention}   
 		\label{force}
 	\end{figure}

A torque-vectoring controller was developed to ensure lateral/yaw stability of the vehicle. The primary objective of this controller is to provide a safe driving experience in the case of unexpected driving condition e.g. low surface friction condition. The desired C.G forces ($F_{des}$) in \cref{structure} are represented as:
  \begin{equation}
F_{des}=[F_x^*, F_y^*, G_z^*]^T
\end{equation}  
   where $F_x^*$,$F_y^*$ and $G_z^*$ are the desired C.G longitudinal and lateral forces and yaw moment, respectively. The actual C.G forces of the vehicle are:
\begin{equation}
F=[F_x, F_y, G_z]^T
\end{equation}  
where $F_x$, $F_y$ and $G_z$ are the actual C.G longitudinal and lateral forces and yaw moment of the vehicle, respectively. Each C.G force components is a function of longitudinal and lateral tire forces (see \cref{force}) as: 
\begin{align}
F_x&=F_x(f_{x1},\dots, f_{x4}, f_{y1},\dots,f_{y4}) \label{1}\\
F_y&=F_y(f_{x1},\dots, f_{x4}, f_{y1},\dots,f_{y4}) \label{2}\\
G_z&=G_z(f_{x1},\dots, f_{x4}, f_{y1},\dots,f_{y4}) \label{3} 
\end{align} 

where, $f_{xi}$ and $f_{yi}$,$(i=1,\dots,4)$  are longitudinal and lateral tire forces on each wheel of the vehicle, respectively. Equations \eqref{1}-\eqref{3} reveal a possible way to obtain the desired C.G forces by controlling the tire forces. The tire force vector can be represented as:
\begin{equation}
	f= [f_{x1},f_{x2},f_{x3},f_{x4},f_{y1},f_{y2},f_{y3},f_{y4}]^T
\end{equation} 
The corresponding adjusted C.G forces to minimize the error between the actual forces $F(f)$ and desired forces $F_{des}$, can be written as:
\begin{equation}
F(f+\delta_F)\approx F(f)+\nabla F(f)\delta_F
\label{T1}
\end{equation}   

where, $\nabla F(f)$ is the Jacobian matrix and $\delta_{F}$ is the vector containing the control actions needed to minimise the error between the actual and desired forces which is:
\begin{equation}
	\delta_F= [\delta_{f_{x1}},\delta_{f_{x2}},\delta_{f_{x3}},\delta_{f_{x4}},\delta_{f_{y1}},\delta_{f_{y2}},\delta_{f_{y3}},\delta_{f_{y4}}]^T
	\label{totalforce}
\end{equation} 
It is noted that in this work in order to derive torque vectoring control actions, only $\delta_{f_{xi}}$is considered for controller designing procedure and the effect of lateral control actions are neglected:
\begin{equation}
	\delta f_{yi} = 0,\hspace{0.5cm} i=1,\dots,4
	\label{y0}
\end{equation} 
To formulate the required control action $\delta_{F}$, the error between the desired C.G forces at $t + \Delta t$ and actual C.G forces at $t$ is described:
\begin{equation}
E = F_{des} - F(f) = [E_x, E_y, E_z]^T
\label{e0}
\end{equation} 
Then the control actions $\delta_{F}$ for time $t+\Delta t$ can be calculated from $F_{des} - F(f+\delta_{F})$, where $F(f+\delta_{F})$ is the C.G forces at $t + \Delta t$. Then using \eqref{T1}, \eqref{y0} and \eqref{e0}, one has:
\begin{equation}
	F_{des} - F(f+\delta_{F}) \approx F_{des}-F(f)-\nabla F(f)\delta F=E-\nabla F(f)\delta F
\end{equation} 
where 
\begin{equation}
\nabla F = \begin{bmatrix}
	 \frac{\delta F_x(f)}{\delta F} \\
	\frac{\delta F_y(f)}{\delta F}\\
	\frac{\delta G_z(f)}{\delta F}   
\end{bmatrix}=\begin{bmatrix}
		\frac{\delta F_x}{\delta f_{x1}} & \frac{\delta F_x}{\delta f_{x2}} & \frac{\delta F_x}{\delta f_{x3}} & \frac{\delta F_x}{\delta f_{x4}}\\
			\frac{\delta F_y}{\delta f_{x1}} & \frac{\delta F_y}{\delta f_{x2}} & \frac{\delta F_y}{\delta f_{x3}} & \frac{\delta F_y}{\delta f_{x4}}\\
		\frac{\delta G_z}{\delta f_{x1}} & \frac{\delta G_z}{\delta f_{x2}} & \frac{\delta G_z}{\delta f_{x3}} & \frac{\delta G_z}{\delta f_{x4}}
\end{bmatrix}
\end{equation} 
Then an objective function consists of weighted combination of error between the actual and desired vehicle C.G forces and control action must be defined. The mathematical representation of objective function is:     
\begin{equation}
\begin{gathered}
P = \frac{1}{2}[(E-\nabla F \delta_F)^TW_E(E-\nabla F \delta_F)+\\\delta_F^TW_{df}\delta_F]
\end{gathered}
\label{index}
\end{equation}
where $W_E \in \mathbb{R}^{3x3}$ is the weight on longitudinal, lateral and yaw moment error, and $W_{df} \in \mathbb{R}^{4x4}$ is the weight on control action. Since \eqref{index} is a quadratic form with respect to the tire force adjustment $\delta F$, the necessary condition of the solution is given by solving the equation
\begin{equation}
\frac{\partial P}{\partial \delta F}=0 
\end{equation} 
As a result of minimization of the objective function, the control action required to stabilize the vehicle is \cite{Fallah2013}:
   \begin{equation}
\begin{gathered}
\delta F=[W_{df}+(\nabla F(f)^TW_E)\nabla F(f)]^{-1}
(\nabla F(f)^T(W_E)E)
\end{gathered}  
\end{equation}  
 The applied corrective torque on each wheel is $\delta T = R_{eff}\times\delta F$ where $R_{eff}$ is the effective wheel radius. It is noted that in this paper, there is no adjustment on lateral tire forces and only longitudinal corrective actions are calculated from the controller design. Subsequently, only the weights on the longitudinal corrective actions are tuned using reinforcement learning algorithm, while keeping $W_E$ as fixed values.
\section{Reinforcement Learning Algorithm} 
      \label{RL}
In this section an overview of reinforcement learning framework followed by DDPG algorithm used for parameter tuning of torque-vectoring controller is presented.
\subsection{Reinforcement learning overview}
To implement a Reinforcement Learning (RL) based controller, we consider a standard RL setup for continuous control in which an agent interacts with the environment, aiming to learn from its own actions. The formulation of reinforcement learning is based on a Markov Decision Process. At each time step $t, t=0,1,...,T$, the agent receives an observations $s_t$, takes an action $a_t$ from a set of possible action $\mathcal{A}$. As a consequence of the actions, the agent receives a reward $r_t$ and observes a new state $s_{t+1}$ \cite{Kuutti2019}\cite{Lillicrap2016}. The goal in reinforcement learning is to learn a policy $\pi$, which maps states to actions: $\pi$: $\mathcal{S}\rightarrow\mathcal{A}$ resulting in maximizing the expected cumulative discounted reward $R=\mathbb{E}[\sum_{t=0}^{t=T}\gamma^tr(s_t,a_t)]$ with discounting factor $\gamma\in[0,1]$ and $\mathbb{E}$ denoting the expectation of the probability. The state-action value ($Q$ value) at time step $t$ represents the expected cumulative discounted reward from time $t$. Reinforcement learning problem is solved using Bellman's principle of optimality. That is, if the optimal value of the state-action $Q^*(s_{t+1},a_{t+1})$ for the next time step is known, then the optimal state-action value for the current time step can be calculated by maximizing $r(s_t,a_t)+Q^*(s_{t+1},a_{t+1})$.\\    
In this paper, the actions of the reinforcement learning framework are the control action weighting parameters $W_{df}$. The observation for reinforcement learning are the errors between the actual and desired forces at the C.G of the vehicle as well as the vehicle states $s_t=[e_x,e_y,e_{Mz}, \dot{x},\dot{y},\psi,\dot{\psi},X,Y]$.\\ To encourage the agent to tune the weighting parameters, a reward function must be appropriately defined by the user. In our model the objective is to find a selection of weighting parameters to enforce the vehicle to follow the reference signals while maintaining the stability. Therefore, a reward function must be designed such that the weighting matrix $W_{df}$ generated from reinforcement learning makes the reward function and the objective function in \eqref{index} to be as similar as possible. Hence, the error between the actual and reference C.G forces that are used in the cost function \eqref{index} are chosen for the reward function in order to find a suitable selection of weighting matrix for torque vectoring controller. It is noted that proper selection of a reward function determines the behavior of controller, and thus affects the stability of the vehicle. Therefore, in order to ensure the reliability of the torque vectoring controller a well-defined reward function $r_t$ provided at every time step $t$ is introduced:
   \begin{equation}
   r_t=-(10e_x^2+5e_y^2+5e_{Mz}^2)\times0.01+6M_t + N_t
   \label{reward}
   \end{equation} 
The first term in the reward function encourages the agent to minimize the errors, while the logic conditions encourage the agent to keep the error below some threshold. The two logics in \eqref{reward} are: $M_t=0$ if the simulation is terminated, otherwise $M_t=1$. $N_t=1$ if the components of error is $e_x,e_y<\SI{200}{\newton}$ , otherwise $N_t=0$. In this way a large positive reward is applied when the agent is close to its ideal conditions which is when the errors of C.G are small. On the other hand, a negative reward is applied when the vehicle fails to maintain the stability, which discourages the agent from losing the directionality of the vehicle. It is noted that the simulation is terminated when the sideslip angle of the vehicle is greater than $\SI{8}{\degree}$. Moreover, the action signal takes the values between $40$ and $1e^3$ as lower and upper bound for all parameters in $W_{df}$.   
\begin{table}[t!]
	\caption{DDPG parameter values}
	\label{table_example}
	\begin{center}
		\begin{tabular}{c||c}
			\hline
			\hline
			Reward discount factor & 0.99\\
			\hline
			Critic learning rate & $1e^{-3}$\\
			\hline
			Actor learning rate & $1e^{-4}$\\
			\hline
			Target network update & $1e^{-3}$\\
			\hline
			Mini-batch size & 70\\
			\hline
			Variance decay rate & $1e^{-3}$\\
			\hline
			Variance & $30$\\
			\hline
			Sample time & 0.5\\
			\hline
			Number of neurons & 100\\
				\hline
			Experience buffer length & $1e^6$\\
			\hline
			\hline
		\end{tabular}
	\end{center}
\end{table}
\subsection{Deep Deterministic Policy Gradient}
DDPG is an evolution of Deterministic Policy Gradient \cite{Silver2014} algorithm. It is in the family of actor-critic network \cite{S.Sutton2015}, model-free, off-policy algorithms which utilize Deep Neural Networks as function approximators. DDPG allows to learn policies in high-dimensional, continuous state and action spaces. The DDPG used in this paper, is inspired from \cite{Lillicrap2016}. A brief description of the theory will be discussed in this section, however, a keen reader is encouraged to read the original paper. The DDPG algorithm uses two deep neural networks: actor and critic network. The actor network is responsible for state-action mapping through the deterministic $\pi(s_t|\theta^\pi)$ where $\theta^\pi$ represents the actor neural network weight parameters, and critic network is for $Q$-value function $Q(s_t,a_t)$ 
   \begin{equation}
   \small
Q^\pi(s_t,a_t) = \mathbb{E}_{r_t, s_{t+1}\sim{E}}[r(s_t,a_t)+\gamma(Q^\pi(s_{t+1},\pi(s_{t+1}))]
\end{equation} 
The action value function $Q$ is approximated using DNN with net weights parameters $\theta^Q$. For learning the Q-value, Bellman's principle of optimality is used to minimize the root mean squared loss
   \begin{equation}
L(\theta^Q) = \mathbb{E}_{s_t\sim\rho^\beta, a_t\sim\beta,r_t\sim{E}}[(Q(s_t,a_t|\theta^Q)-y_t)^2]
\end{equation} 
where 
\begin{equation}
 y_t=r(s_t,a_t)+\gamma{Q(s_{t+1},\pi(s_{t+1}|\theta^Q))}
 \end{equation}
The actor policy $\pi(s_t|\theta^\pi)$ is updated using:
\begin{equation}
\small
\nabla_{\theta^\pi}\approx \mathbb{E}_{s_t\sim\rho^\beta}[\nabla_{a}Q(s,a|\theta^Q)|_{s=s_t,a=\pi(s_t)}\nabla_{\theta_\pi}\pi(s|\theta^\pi)|_{s=s_t}]
\end{equation}
	\begin{figure}[t!]
		\hspace*{-0.2in}
		\centering
		\includegraphics[width=0.95\linewidth]{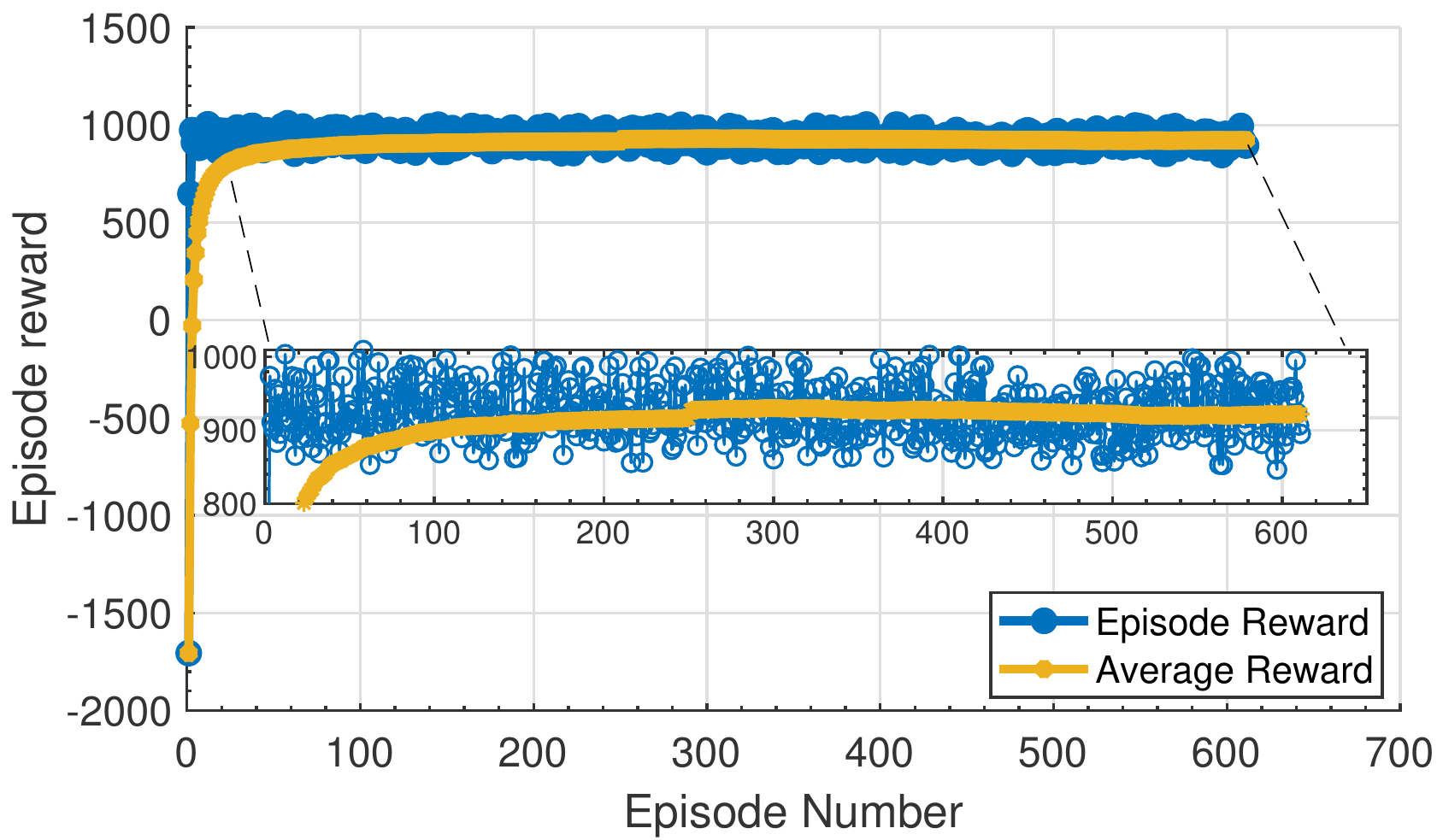}
		\caption {Episode reward for training the torque vectoring parameters}   
		\label{Reward} 
	\end{figure} 
 which is the policy gradient. For learning the policy $\pi$ gradient ascent is performed with respect to the policy parameter $\theta^\pi$ to maximize the Q-value. In DDPG the target networks are used to stabilize the training \cite{Mnih2015}. Gaussian noise is used for action exploration \cite{UHLENBECK1930}. Experience replay is utilized for stability \cite{Mnih2015}. Mini-batch gradient descent is used \cite{Lillicrap2016}. The network parameters were empirically tuned, and final hyperparameters can be found in \cref{table_example}. 
 	\begin{figure}[t!] 
 	\hspace*{0.1in}
 	\centering
 	\includegraphics[width=0.9\linewidth]{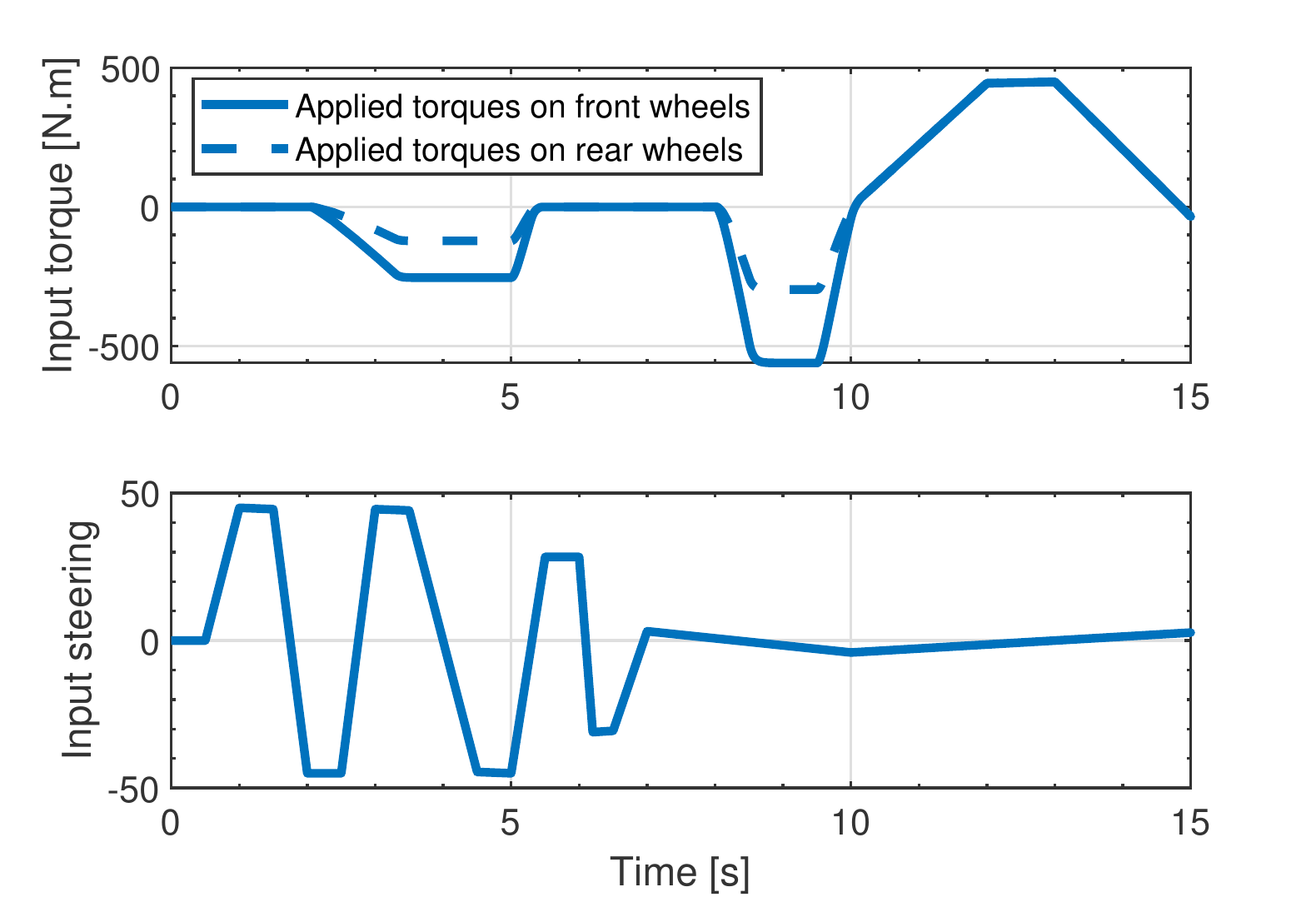}
 	\caption {\textbf{top figure}: Applied torque on each wheel, \textbf{bottom figure}: Input steering}   
 	\label{inputs-dr}
 \end{figure} 
Both the actor and critic networks are neural nets with 2 hidden layers. The hidden neurons all use Relu activation, actor output uses tanh activation, whilst critic output uses linear activation.  
\section{Simulation results}
\label{S-results}
\begin{figure}[t!] 
		\hspace*{-0.1in}
		\centering
		\includegraphics[width=0.9\linewidth]{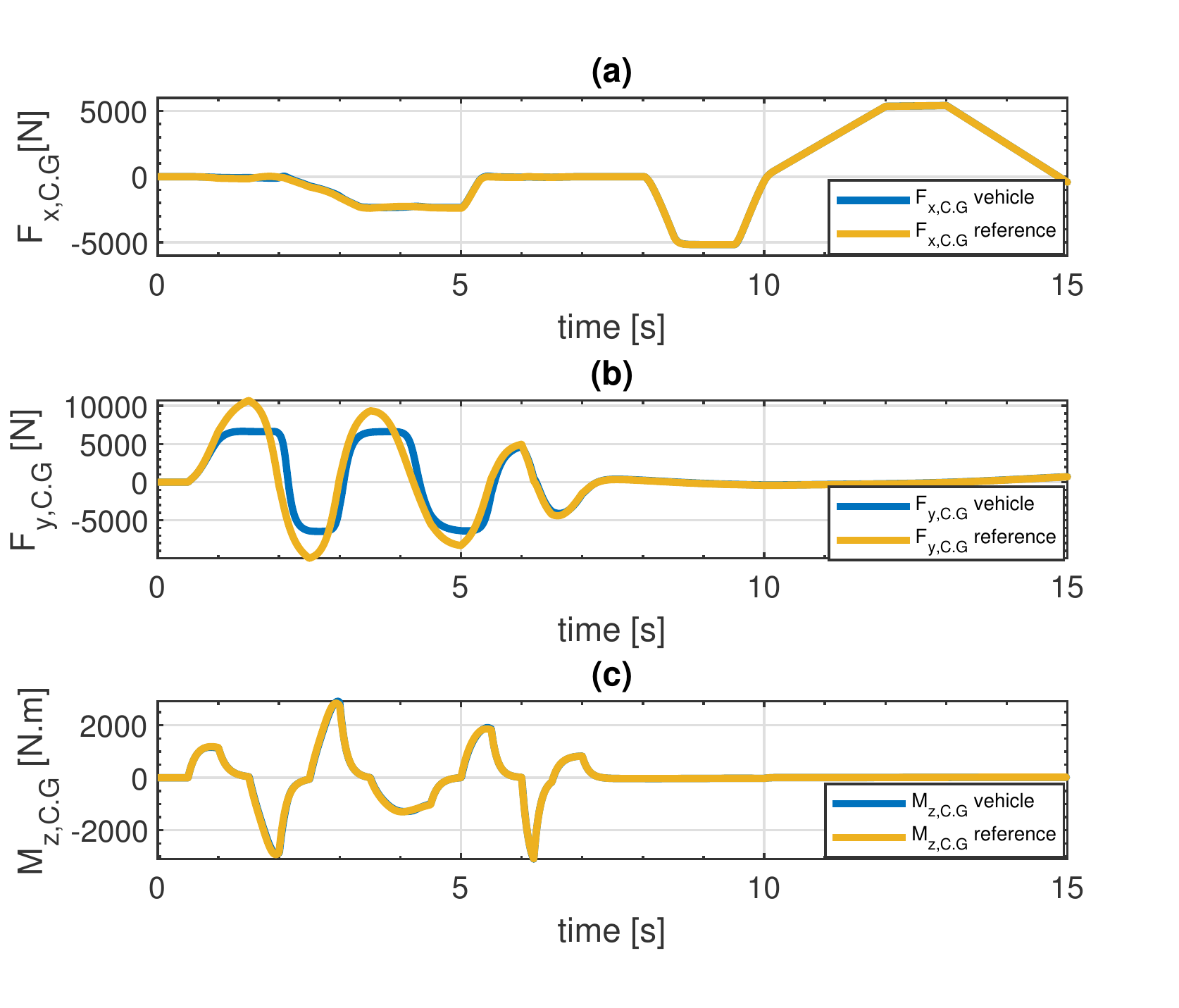}
		\caption {Longitudinal, lateral forces and yaw moment around C.G The yellow lines represent the actual vehicle behavior, and blue line represent the reference forces estimated from CIM block}   
		\label{forces}
	\end{figure}

 The DDPG agent was trained for 650 episode with $\SI{15}{\second}$ episode length. At the start of each training episode, the environment parameters were varied by choosing initial velocity randomly between $\SI{80}{\kilo\meter\per\hour}$ to $\SI{130}{\kilo\meter\per\hour}$ and friction value ranging from 0.4 to 0.6. As can be seen from the episode rewards (see \cref{Reward}), the agent quickly converges to an optimal policy, although slight improvements in the average reward can still be seen in the later episodes. Once the training phase is completed, performance of the DDPG network is validated in various driving scenarios. All simulation experiments presented in this paper were performed on a four wheel vehicle model with pacejka tire model described in \cref{vehicle-model}. For the simulation purposes the driver's input torque and steering wheel angle from reference generator are shown in \cref{inputs-dr}. The parameters used to design vehicle model are tabulated in \cref{table_vehicle}. In this section, we first demonstrated the results of the algorithm in scenarios that DDPG was trained in and compared the simulation results with genetic algorithm and manual tuning of the weighting matrix $W_{df}$. It is noted that the error signals used in the reward function \eqref{reward} are chosen for designing cost function in genetic algorithm in order to have a fair comparison between two strategies. For further validation, the effectiveness of the DDPG algorithm is investigated for the conditions that reinforcement learning was not trained in, under different input steering angle. This analysis aims to investigate the generalization capability of the DDPG tuning strategy, by testing its performance in scenarios beyond the training environment. The first scenario that is chosen to demonstrate the performance of DDPG algorithm, is under conditions where the surface friction is 0.4 and initial velocity is $\SI{100}{\kilo\metre\per\hour}$. 
  	\begin{figure}[t!] 
 	\hspace*{-0.2in}
 	\centering
 	\includegraphics[width=0.8\linewidth]{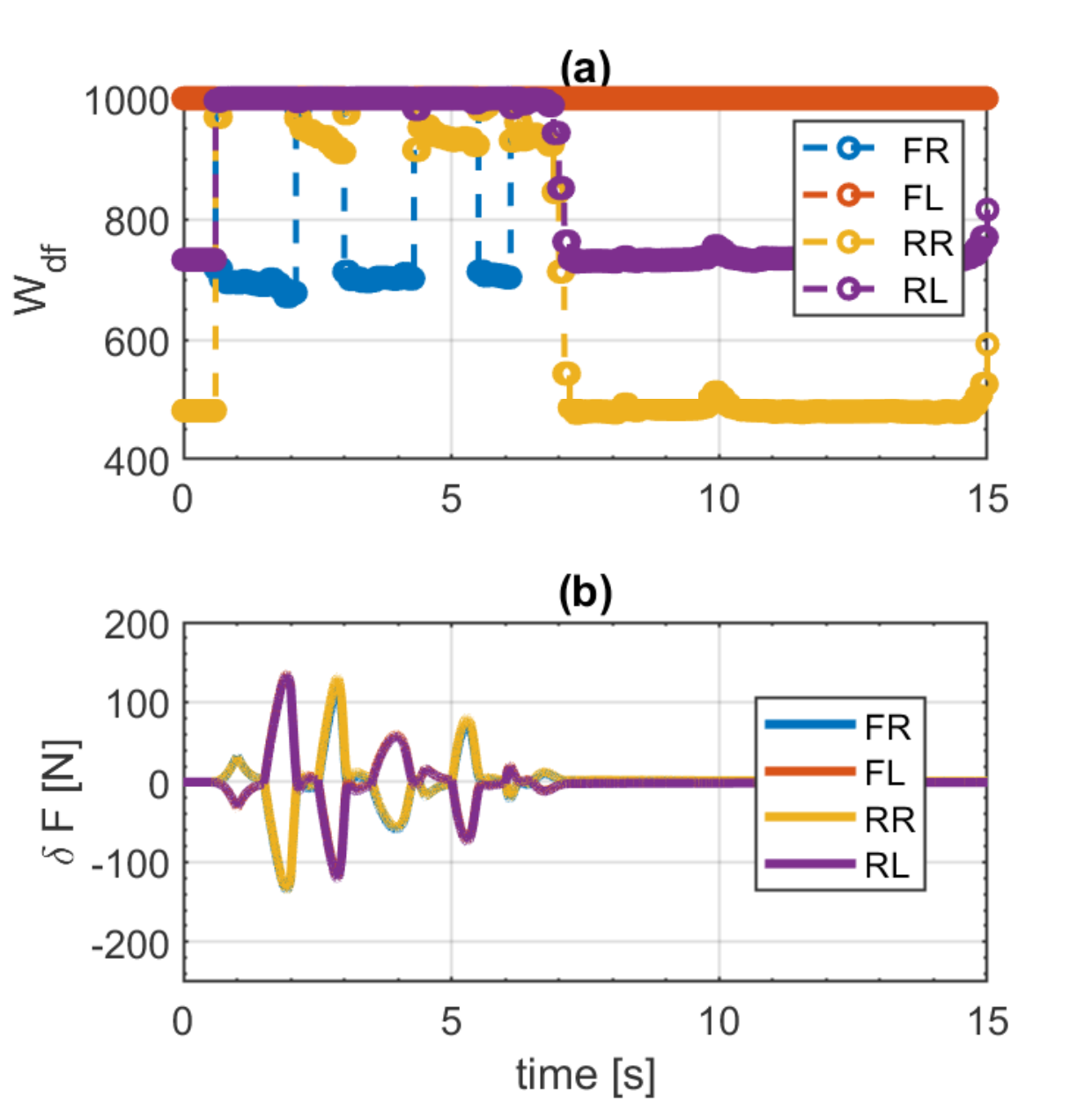}
 	\caption {\textbf{(a)}: Generated weighting matrix $W_{df}$, \textbf{(b)}: Applied torque on each wheels}   
 	\label{inputs}
 \end{figure} 
 It is noted that every iteration of the algorithm is actively trying to maximise the reward function by reducing the error signals presented in \eqref{reward}, and at the same time maintaining a smooth and low control effort from torque vectoring controller by generating an optimal set of weighting matrix from reinforcement learning. This can be observed in \cref{inputs} where the top figure shows the weighting matrix generated from the reinforcement learning algorithm while the bottom figure shows the control action from torque vectoring controller. The generated adjustment torque $\delta F$ on each wheel shows a symmetric and smooth evolution as a consequence of weighting matrix $W_{df}$. It is noted that the course of $W_{df}$ on the front left (FL) wheel is constant over the entire simulation compared to the other weights on each wheel. This behavior indicates that reinforcement learning found some semi-trivial solution to the problem for the entire operating regions, demonstrating the generalization capabilities of the trained model which will be clear later in this section.\\
 \begin{table}[b!]
 	\caption{Design parameters}
 	\label{table_vehicle}
 	\begin{center}
 		\begin{tabular}{c||c}
 			\hline
 			\hline
 			Vehicle mass  & 1360 $kg$\\
 			\hline
 			Yaw moment of inertia & 2050 $kgm^2$\\
 			\hline
 			Effective wheel radius & 0.3 $m$\\
 			\hline
 			Front distance from vehicle C.G & 1.43 $m$\\
 			\hline
 			Rear distance from vehicle C.G & 1.21 $m$\\
 			\hline
 			Front cornering stiffness & $\SI{16000}{\newton\per\radian}$\\
 			\hline
 			Rear cornering stiffness & $\SI{16000}{\newton\per\radian}$\\
 			\hline
 			$W_E$& $diag(0.4,0.02,1500)$\\
 			\hline
 			\hline
 		\end{tabular}
 	\end{center}
 \end{table}
\cref{forces}, compares the actual C.G forces obtained from the vehicle model with those estimated by CIM block. As can be seen, the generated C.G forces are able to track the reference values, resulting in stabilizing the vehicle under friction surface $\mu = 0.4$. However, the small discrepancy can be seen in the second figure of the subplot. This is due to the fact that in the formulation presented in \cref{control}, since there is no direct control of the lateral tire forces, the magnitude of the error in lateral C.G force is relatively large, hence it would be difficult to fully minimize the lateral C.G error under this condition.\par 
\begin{figure}[t!] 
		\hspace*{-0.1in}
		\centering
		\includegraphics[width=0.9\linewidth]{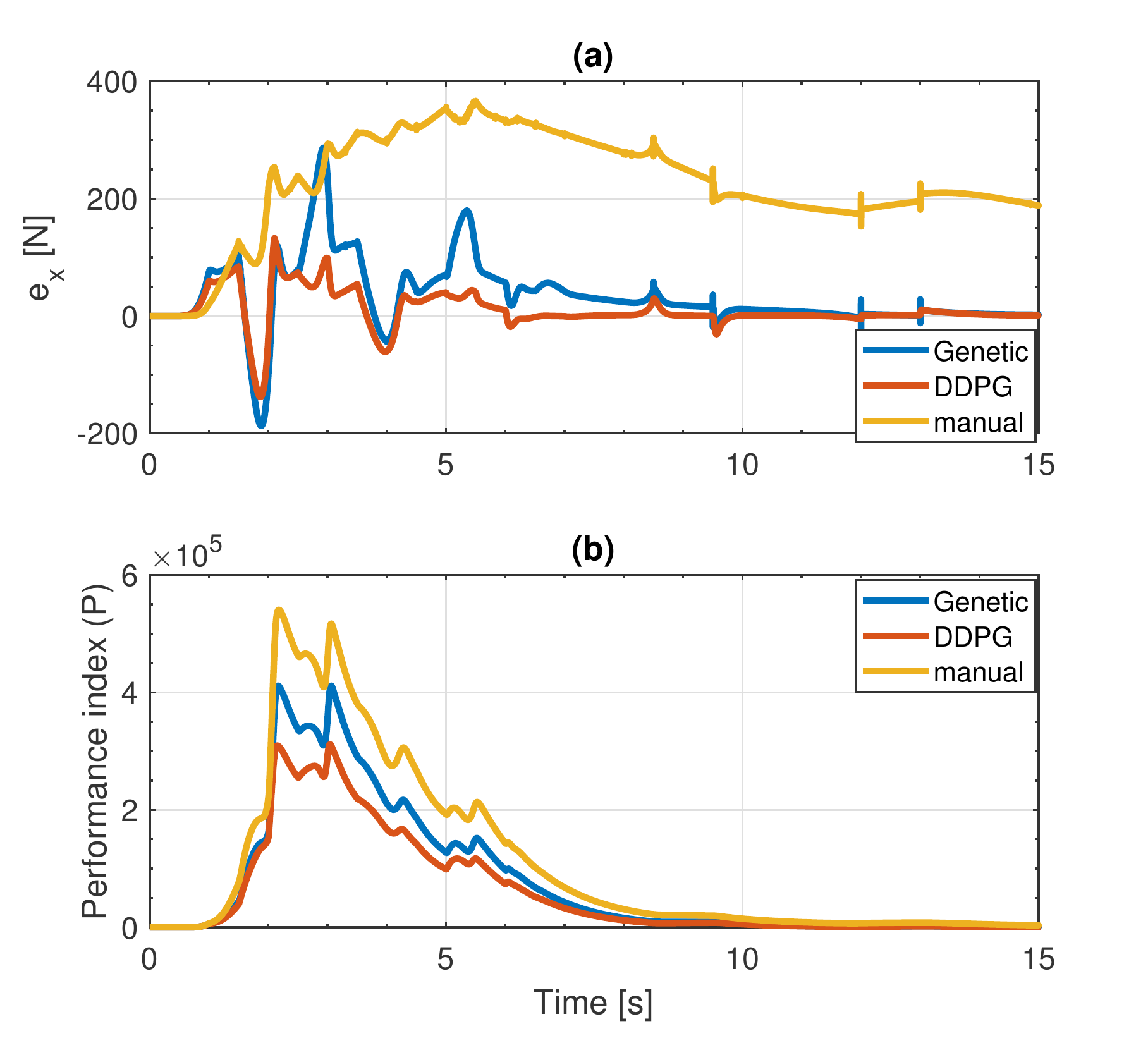}
		\caption {\textbf{(a)}: Longitudinal C.G force errors under $\mu = 0.4$, \textbf{(b)}: Performance index results, \textbf{note}: Blue line represents genetic algorithm tuning results, yellow line represents manual tuning, and orange line is DDPG algorithm results}   
		\label{efx}
	\end{figure}

To validate the effectiveness of the DDPG algorithm, the result of the longitudinal C.G error is compared with genetic algorithm and trial-and-error approach (manual tuning). It is noted that, the manual tuning of $W_{df}$ was selected based on our previous work in \cite{Taherian2019} where its corresponding values resulted in maintaining the vehicle stability with $W_{df} = diag(1e^2, 1e^2, 1e^2, 1e^2)$. \cref{efx} shows the comparison between parameter tuning using DDPG algorithm, genetic algorithm and manual tuning of torque-vectoring controller.  \cref{efx}(a) demonstrates the evolution of error in longitudinal C.G forces for all tuning approaches. In this analysis the value of errors are such that vehicle can successfully maintain the stability. However, evolution of the error for DDPG algorithm outperforms the rest of the approaches with the absolute value of the maximum error of $\SI{137}{\newton}$, whereas in genetic algorithm and manual tuning this value reaches to $\SI{205}{\newton}$ and $\SI{366}{\newton}$ respectively.
Moreover, unlike the manual tuning, the value of the error for both DDPG and genetic algorithm is able to return to near-zero, however the magnitude of the error for DDPG is smaller compared to genetic algorithm which shows the superiority of DDPG strategy over the rest of the approaches. This confirms that DDPG algorithm is a reliable technique to obtain better understanding of controller parameters and the tuning, particularly in the nonlinear region of operation of the vehicle. Note that, in this paper, only error on the longitudinal C.G force is analyzed, since only the longitudinal corrective action is considered to be controlled in order to stabilize the vehicle. \cref{efx}(b) demonstrates the outcome of the objective function \eqref{index} to show the optimal solution of the torque vectoring controller as a result of parameter tuning. As can be seen, DDPG is able to find the best solution by reducing the performance index to the minimum level and converges to the near-zero value around $\SI{8}{\second}$ of the simulation. This optimal value shows better result compared to genetic algorithm and almost double improvement compared to manual tuning of the weighting matrix.\\
Parametric analysis has been carried out to verify the effectiveness of DDPG algorithm. In this analysis, the absolute value of the maximum error of longitudinal C.G error force for DDPG algorithm, genetic algorithm and manual tuning of torque-vectoring controller is presented (see \cref{table_simresults}).  First, the analysis focuses on comparing the DDPG algorithm with other approaches, under different friction coefficient of the road and the same initial velocity of $\SI{100}{\kilo\meter\per\hour}$. In this analysis, the value of error is increasing gradually for all cases as the road friction decreases. However, for all friction values, DDPG reduces the maximum absolute errors by more than 50\% which shows improvement compared to genetic algorithm and manual tuning.
	\begin{table*}[t!]
	\caption{Longitudinal C.G error for various driving scenarios}
	\label{table_simresults}
	\centering
	\begin{tabular}{ m{3cm} m{1cm}  m{1cm}  m{1cm} m{1cm} m{1cm} m{1cm} m{2cm}}
		\hline
		\hline
		\textbf{$\mu$} & \textbf{0.4} & \textbf{0.45} & \textbf{0.5} & \textbf{0.55}& \textbf{0.6}& \textbf{0.65}& \textbf{0.7}
		\\
		\hline\\
		DDPG & 132 N & 115 N & 77 N & 54 N & 43 N & 32 N & 18 N\\
		Genetic algorithm & 205 N & 180 N & 163 N & 105 N & 86 N & 65 N & 32 N\\
		Manual tuning & 366 N & 289 N & 223 N & 170 N & 127 N & 95 N & 70 N \\
		\hline
		\hline
		\textbf{$V_x (km/h)$} & \textbf{70} & \textbf{80} & \textbf{90} & \textbf{100}& \textbf{110}& \textbf{120}& \textbf{130}
		\\
		\hline\\
		DDPG & 17 N &33 N & 77 N & 132 N & 182 N & 202 N & 214 N\\
		Genetic algorithm & 40 N &73 N & 161 N & 205 N & 287 N & 322 N & 390 N\\
		Manual tuning & 71 N & 105 N & 212 N & 366 N & 487 N & 498 N & 515 N\\
		\hline
		\hline
	\end{tabular}
\end{table*}
	\begin{figure}[t!] 
		\hspace*{-0.1in}
		\centering
		\includegraphics[width=0.86\linewidth]{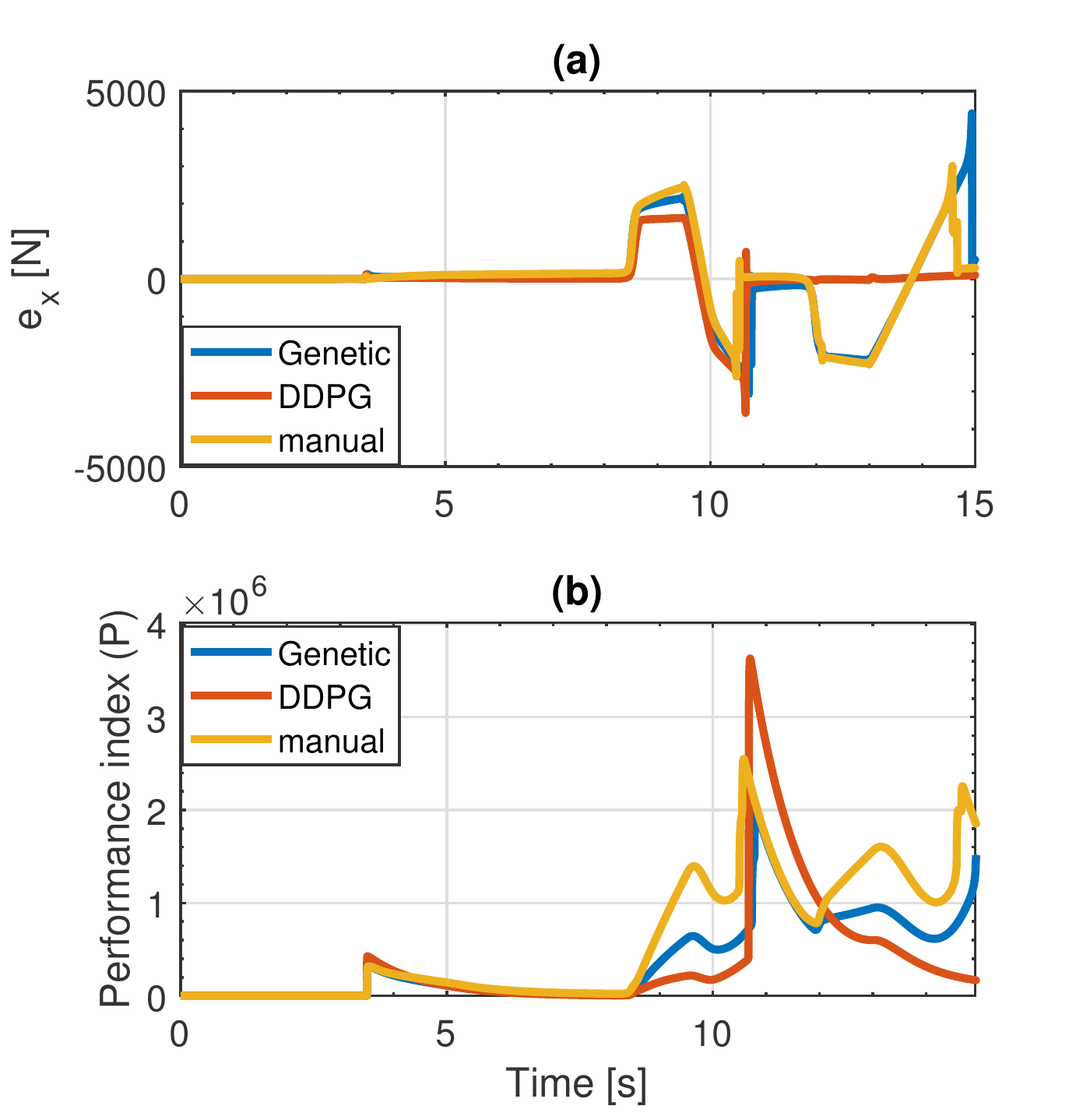}
		\caption {\textbf{(a)}: Longitudinal C.G force errors under $\mu = 0.3$, \textbf{(b)}: Performance index results}   
		\label{error-diff}
	\end{figure}  
 This analysis shows the effectiveness of DDPG algorithm on the performance of the control structure in adapting of the system to the changes of the environment condition. The second analysis has been conducted to evaluate the proposed framework under different velocities with a fixed friction surface condition ($\mu=0.4$). It can be seen that as the velocity of the vehicle increases, the absolute value of the maximum error increases for both cases. This analysis confirms the ability of the DDPG algorithm to improve the adaptive capability of the controller under different velocities and low surface friction condition.\\
Finally to validate generalization capability of the DDPG algorithm, simulation results have been conducted under the environment that DDPG was not trained in. Therefore, friction surface of $\mu=0.3$ is chosen as an environment that was not considered during training. Initial velocity of the vehicle is chosen as $\SI{80}{\kilo\meter\per\hour}$ which is a reasonable representation of high-speed driving on an icy/snow surface. Additionally, a $\SI{50}{\degree}$ step steer manoeuvre is applied as an input steering angle ($\delta_f$) to the system in order to investigate the generalization capability of DDPG algorithm under the unforeseen conditions. This can be seen in \cref{error-diff} where in \cref{error-diff}(a) the longitudinal C.G error shows larger evolution during braking and acceleration for the duration of 8 to 11s for all tuning approaches due to sudden input driving torque applied to the controller. However, as can be seen the error for DDPG algorithm converges to zero which indicates reasonably good tracking of this algorithm even in the environment that it was not trained in. On the other hand, genetic algorithm and manual tuning show larger evolution of the error and fail to converge to zero with slightly better tracking capability of genetic algorithm. \cref{error-diff}(b) demonstrates the result of the cost function \eqref{index} in order to show the optimal solution of the torque vectoring controller for all approaches. As can be seen DDPG offers a better solution compared to other methods despite a relatively large peak around 11s. The reason of this is due to a large $W_{df}$ (see \cref{inputs}(a)) applied to the controller to maintain the vehicle stability, which ultimately minimised the performance index to near-zero value and find an optimal solution. On the other hand the value of performance index for genetic algorithm and manual tuning observed an increasing trend with lack of converging to an optimal solution. The results show the superiority of the proposed DDPG approach to parameter tuning over the manual tuning and genetic algorithm baselines. More importantly, by carrying out tests with different road conditions, vehicle velocities, and steering inputs from those seen during training, the results demonstrate that the DDPG has learned a general approach to tune the weighting parameters of the torque vectoring controller which can generalize to new scenarios and environments.
\section{CONCLUSIONS}
\label{conclusion}
This paper has investigated the use of DDPG algorithm for automatically tuning a torque-vectoring controller under a wide range of different vehicle velocities and friction surface conditions. The proposed control framework consists of \textit{(i)} reinforcement learning method as an auto-tuning algorithm, \textit{(ii)} a torque-vectoring controller to ensure lateral-yaw stability of the vehicle. The closed-loop scheme was implemented on a four wheels vehicle model with nonlinear tire characteristics, and the numerical results indicate the benefits of the reinforcement learning on tuning the torque-vectoring parameters. The DDPG has multiple advantages over genetic algorithm and manual tuning as it can find better weighting parameters of the controller, as well as tuning the parameters in an online manner based on the current states of the vehicle. The results demonstrated the effectiveness of the DDPG algorithm as an auto-tuning method over wide range of operating points, resulting in significant reduction of Longitudinal C.G errors compared to other tuning approaches. The overall performance of the vehicle indicates the accurate tracking of references as well as maintaining the stability of the vehicle using DDPG which shows the efficacy of this algorithm. While this work evaluated the adaptive tuning of torque vectoring parameters via $W_{df}$, this framework is quite general and could be adapted to different controllers and even different domains.





\bibliographystyle{IEEEtran}
\bibliography{References}


\end{document}